\documentclass[sigconf]{acmart}
\usepackage{amsmath}
\usepackage{pifont}
\usepackage{enumitem}

\usepackage{subcaption}

\definecolor{chgup}{HTML}{E8F5E9}     
\definecolor{chgnew}{HTML}{E3F2FD}    
\definecolor{chgsame}{HTML}{F5F5F5}   
\definecolor{accepted}{HTML}{F3E5F5}  
\definecolor{dropped}{HTML}{FFEBEE} 

\usepackage{listings}

\usepackage{multirow} 

\usepackage{mathtools}
\usepackage{enumitem}

\usepackage{ifthen}
\usepackage{fontawesome5}
\usepackage{color}

\usepackage[utf8]{inputenc}
\usepackage[linesnumbered, ruled, vlined]{algorithm2e}

\newcommand{\circled}[1]{%
  \textcircled{\raisebox{-0.15ex}{\scriptsize\textbf{#1}}}%
}

\newboolean{showcomments}
\setboolean{showcomments}{true} 
\ifthenelse{\boolean{showcomments}}
{\newcommand{\nb}[2]{
    \fcolorbox{gray}{yellow}{\bfseries\sffamily\scriptsize#1}
    {$\blacktriangleright$#2$\blacktriangleleft$}
  }
  
}
{\newcommand{\nb}[2]{}
  
}

\usepackage{xcolor}
\usepackage{listings}
\usepackage{makecell}
\definecolor{archbg}{HTML}{FFF8E1}  
\definecolor{relbg}{HTML}{E3F2FD}   
\definecolor{attrbg}{HTML}{F3E5F5}  
\definecolor{archframe}{HTML}{FFC107}
\definecolor{relframe}{HTML}{42A5F5}
\definecolor{attrframe}{HTML}{AB47BC}
\definecolor{hdr}{HTML}{37474F}     
\definecolor{fmt}{HTML}{1B5E20}     
\definecolor{blk}{HTML}{004D40}     

\usepackage{pifont}

\newcommand{\xmark}{\ding{55}} 

\newcolumntype{Y}{>{\raggedright\arraybackslash}X}

\copyrightyear{2026}
\acmYear{2026}
\setcopyright{cc}
\setcctype{by}
\acmConference[MODELS 2026]{ACM/IEEE 29th International Conference on Model Driven Engineering Languages and Systems}{October 04--09, 2026}{Málaga, Spain}
\acmBooktitle{ACM/IEEE 29th International Conference on Model Driven Engineering Languages and Systems (MODELS 2026), October 04--09, 2026, Málaga, Spain}
\acmDOI{10.1145/3822455.3830329}
\acmISBN{979-8-4007-2809-9/2026/10}

    
\begin{document}
\title{Towards Automated Domain Model Extraction from Source Code using Heuristics and Open-Source LLMs}




\author{Alessandra Mancas}
\orcid{0009-0008-2272-6693}
\affiliation{%
  \institution{Université de Montréal}
  \department{DIRO}
  \city{Montreal}
  \country{Canada}
}
\email{alessandra.thais.mancas@umontreal.ca}

\author{Mounir Ammam}
\orcid{0009-0000-5190-0795}
\affiliation{%
  \institution{Université de Montréal}
  \department{DIRO}
  \city{Montreal}
  \country{Canada}
}
\email{mounir.ammam@umontreal.ca}

\author{Hyacinth Ali}
\orcid{0000-0003-2287-2646}
\additionalaffiliation{%
\institution{Université de Montréal}
\department{DIRO}
\city{Montreal}
\country{Canada}
}
\affiliation{%
  \institution{University of Alberta}
  \department{Electrical and Computer Engineering}
  \city{Alberta}
  \country{Canada}
}
\email{hcali@ualberta.ca}

\author{Kevin Delcourt}
\orcid{0009-0005-2988-7308}
\affiliation{%
  \institution{Université de Montréal}
  \department{DIRO}
  \city{Montreal}
  \country{Canada}
}
\email{kevin.delcourt@umontreal.ca}

\author{Houari Sahraoui}
\orcid{0000-0001-6304-9926}
\affiliation{%
  \institution{Université de Montréal}
  \department{DIRO}
  \city{Montreal}
  \country{Canada}
}
\email{sahraouh@iro.umontreal.ca}

\begin{abstract}

Large language models (LLMs) have recently shown strong capabilities for code understanding, making them promising for reverse engineering domain models from source code. However, state-of-the-art proprietary LLMs cannot be used in many industrial contexts due to privacy and confidentiality constraints, while compact open-source LLMs that can run locally are limited by their context window and cannot process large code bases directly. 

In this paper, we propose an automated approach to extract domain models from source code using lightweight, locally deployable LLMs. Our method combines structural and semantic heuristics with iterative LLM-based reasoning to overcome context limitations. By progressively analyzing ranked subsets of code elements, the approach identifies domain concepts and refines domain boundaries without requiring full-system context. 

Our approach achieves high F1-scores 
on a dataset of ten projects, each comprising a curated domain model and its corresponding implementation, while remaining fully executable on locally deployable LLMs. This makes it particularly suitable for reverse engineering tasks in privacy-sensitive industrial environments. 



\end{abstract}


\maketitle


\section{Introduction}
\label{sec:intro}

Domain models play a central role in Model-Driven Engineering (MDE), where they are used to represent domain concepts, guide system design \cite{evans2004domain}, and support automated analysis and code generation \cite{broy2013domain}. 
As understanding, maintaining and evolving complex or legacy software remains a significant challenge in the industry \cite{ rajbhoj2025leveraging}, reverse engineering domain models from existing software systems can offer valuable assistance to practitioners \cite{boronat2025mdre}.

Recent advances in generative artificial intelligence have significantly expanded the range of tasks that can be automated in software engineering \cite{amalfitano2026research}. Large language models (LLMs) have demonstrated strong capabilities for code understanding, making them promising candidates for reverse engineering tasks \cite{shi2025llms}, including the extraction of domain models from source code \cite{campanello2025use}. 
While early applications of LLMs to software engineering \cite{idialu2024whodunit} and modeling tasks \cite{kc2024analysis} showed encouraging but limited performance, more recent studies suggest that state-of-the-art proprietary models are approaching human-level performance on such tasks \cite{nguyen2026class}. This is in part due to the fact that these LLMs can process large contexts and perform complex reasoning over code.

However, relying on such models raises ethical concerns in academic, industrial, and governmental contexts. A systematic mapping study by \citet{huang2026ethical}
identifies transparency and privacy as two of the five major concerns associated with their widespread adoption.
Many companies are reluctant or unable to disclose their source code to third-party providers due to security, privacy, and intellectual property constraints. Deploying proprietary LLMs for reverse engineering tasks is often not a viable option in practice.

An alternative is to rely on open-source LLMs that can be executed locally. 
While this addresses privacy and confidentiality concerns, it introduces a different limitation. Compact open-source models typically have restricted context windows, which prevent them from processing large code bases in a single prompt \cite{an2024make}. Consequently, these models cannot directly perform domain model extraction at the scale of real-world systems.

This creates a fundamental gap: powerful LLMs capable of handling large code bases are not deployable in many industrial contexts, while locally executable models lack the capacity to analyze complete systems. There is a need for approaches that enable effective domain model reverse engineering using lightweight, locally deployable LLMs despite their limited context.

In this paper, we propose an automated approach to extract \emph{domain models} from source code using compact open-source language models. 
Our approach combines structural and semantic heuristics with iterative LLM-based reasoning to overcome context limitations. 
Starting from a rough UML class diagram extracted from the code, we compute the semantic similarity between code elements and available project documentation to rank classes according to their likelihood of representing domain concepts. 
We then iteratively process the ranked list of classes, classifying each class as either domain-specific or implementation detail. 
As the process progresses, the language model accumulates contextual knowledge about the domain, enabling it to progressively refine domain boundaries despite operating under limited context.
The identification of attributes and relationships is performed using a similar combination of heuristics and LLM-based reasoning. Beyond extracting domain models, our approach establishes explicit links between model elements and the corresponding implementation artifacts, enabling applications such as model--code traceability and AI-assisted development, both often cited as opportunities in Model-Driven Engineering (MDE) \cite{berardinelli2025model}.
A key advantage of our approach is that it can be executed entirely on compact open-source models, making it suitable for industrial contexts where code confidentiality is critical.

We evaluate our approach on 
ten projects for which both manually defined domain models and corresponding source code implementations are available. 
The results show that our method achieves high precision and recall, while remaining fully deployable in a local setting. 
The main contributions of this paper are:
\begin{itemize}
\item An automated approach for domain model reverse engineering from source code using lightweight, locally deployable language models.
\item A method that combines structural and semantic heuristics with iterative LLM reasoning to overcome context window limitations.
\item A process that establishes explicit links between domain models and implementation code.
\item An empirical evaluation on ten projects conducted on modest hardware. 
\end{itemize}



\section{Background}
\label{sec:background}


\subsection{Domain Modeling}

In Model-Driven Engineering and Requirements Engineering, the domain model is a central artifact representing the key concepts of a problem space \cite{wkasowski2023domain}. 
Domain modeling aims to capture the main concepts, attributes, and relationships that characterize a domain, while excluding implementation details such as operations or interfaces. 
It abstracts real-world entities into structured representations, which can be expressed as graphical models (e.g., UML class diagrams) or textual models (e.g., PlantUML or UMPLE class diagrams~\cite{lethbridge2021umple}). 
As a core artifact in early software design, domain models foster shared understanding between stakeholders and engineers, bridging the gap between business knowledge and system implementation \cite{evans2004domain}.


During requirements elicitation and design, modelers translate textual specifications into UML class diagrams \cite{rumbaugh2005unified}, a process that is time-consuming and requires expertise \cite{chen2023automated}; these models then serve as central artifacts supporting downstream activities such as model transformation, system analysis \cite{dao2025learning}, and facilitate traceability between requirements and implementation.
Domain modeling has been applied to a wide range of industrial applications, such as 
cyber-physical systems \cite{bucaioni2025engineering}, or digital twins \cite{michael2026opportunities}.

\subsection{Model Driven Reverse Engineering}

Reverse engineering 
refers to the process of analyzing software systems to identify their components and recover higher-level representations
from source code \cite{koschke2006architecture}. In the context of MDE, Model-Driven Reverse Engineering specifically focuses on deriving high-level models by leveraging MDE techniques such as metamodeling and model transformations \cite{siala2024model}.
This area addresses long-standing challenges in software engineering practice, including legacy system modernization, maintenance, and evolution, by aiming to produce comprehensible representations such as domain diagrams from source ~\cite{campanello2025use, boronat2025mdre}. 

These challenges are further amplified by the increasing prevalence of LLM-generated code. 
As development teams generate more code with LLMs, it becomes more difficult to maintain a shared understanding of system behavior \cite{storey2026technical}. Consequently, there is a growing need for methods that can abstract and recover the underlying domain concepts of a system, helping to manage this emerging form of cognitive debt \cite{storey2026technical}.

\subsection{Context Windows in Large Language Models}\label{sec:bcg-context-windows}
LLMs
have demonstrated remarkable capabilities across a wide range of natural language processing tasks \cite{bommasani2021opportunities}. 
They are effective at 
generalizing across different modalities \cite{tan2024large}. This makes them key building blocks for intelligent systems addressing various Software Engineering challenges \cite{amalfitano2026research}.

However, these models also face notable limitations. 
In particular, the high computational cost associated with training and deploying large LLMs limits their accessibility for many organizations, while the use of proprietary, closed-source models raises increasing ethical concerns \cite{huang2026ethical}. Smaller LLMs, which can be deployed on cheaper hardware, provide a practical alternative, but are often constrained by limited context windows. As highlighted by \citet{naveed2025comprehensive}, handling long contexts remains a major challenge in current LLM research, and a variety of approaches are being explored to address this limitation.

\paragraph{Approaches for handling large contexts with LLMs.}

A first category of approaches focuses on modifying LLM architectures to natively support longer contexts. 
For example, techniques such as RoPE scaling \cite{pal2023giraffe} extend the usable context length by adapting positional embeddings, enabling models trained on shorter sequences to generalize to longer inputs.
Other works redesign the attention mechanism to reduce the computational cost of standard self-attention \cite{guo2022longt5}. Recent versions of Qwen leverage these techniques to enable relatively small models to handle contexts of up to 1M tokens, as seen with \texttt{Qwen2.5-14B} \cite{ahmed2025qwen}. However, these approaches remain an active area of research and are difficult to deploy on consumer-grade hardware \cite{naveed2025comprehensive} .

A second category of approaches circumvents context window limitations without modifying the underlying model architecture. Retrieval-Augmented Generation (RAG) methods improve prompts with information retrieved from large external corpora \cite{giuffre2024optimizing}. Similarly, memory-augmented LLMs rely on external memory structures to store and retrieve relevant information \cite{zhang2025survey}. These techniques enable models to handle complex tasks despite limited context windows.
Finally, \citet{kyoung2025reasoning} shows that decomposing complex queries into smaller sub-questions improves long-context question answering performance without requiring additional training. This effectively allows models to process complex inputs through multiple focused reasoning steps.

Our work aligns with this latter category. We adopt a multi-step, iterative prompting strategy. However, unlike generic approaches, our method leverages domain-specific knowledge from Model-Driven Engineering to design a structured pipeline tailored to the task, which can lead to more accurate, meaningful and efficient abstractions.

\subsection{Embeddings and Semantic Similarity}

Embeddings are a fundamental representation technique in machine learning, which is extensively used in a wide variety of modern Natural Language Processing (NLP)~\cite{tang2015document}
and generation~\cite{zhang2017adversarial} tasks. They map discrete data, such as words, sentences, or code elements, into a continuous high-dimensional vector representation, where semantic relationships are encoded geometrically. Existing approaches, such as Word2Vec~\cite{ma2015using}, demonstrate that semantic regularities can be captured through vector representations learned from large corpora, which allows meaningful comparisons between textual units~\cite{mikolov2013efficient}. This idea has since been extended to contextualized representations using transformer-based models, which encode richer syntactic and semantic information.

Recently, models, such as Sentence-BERT (SBERT)~\cite{fujishiro2023accuracy}, generate semantically meaningful embeddings for documents, which support efficient comparison using similarity measures such as cosine similarity. 
These embedding-based approaches have become the standard for semantic textual similarity tasks, where the goal is to quantify the degree of semantic equivalence between two pieces of text~\cite{guder2026sentence}. Given two vectors, \emph{a} and \emph{b}, their similarity can be expressed using cosine similarity as:

\begin{equation}
    \label{eq:placeholder_label}
    \text{similarity}(\mathbf{a}, \mathbf{b}) = \frac{\mathbf{a} \cdot \mathbf{b}}{\lVert \mathbf{a} \rVert \, \lVert \mathbf{b} \rVert}
\end{equation}

Cosine similarity is widely adopted as it captures the angular relationship between vectors and is empirically effective for comparing semantic representations across a variety of embedding models~\cite{zhelezniak2019correlation}. In our work, this enables the estimation of semantic similarity between extracted diagram elements (i.e., classes) and the domain description, thereby supporting the extraction of domain diagrams at an appropriate level of abstraction.

\section{Approach}
\label{sec:approach}


\begin{figure}
    \centering
    \fbox{\includegraphics[width=\linewidth]{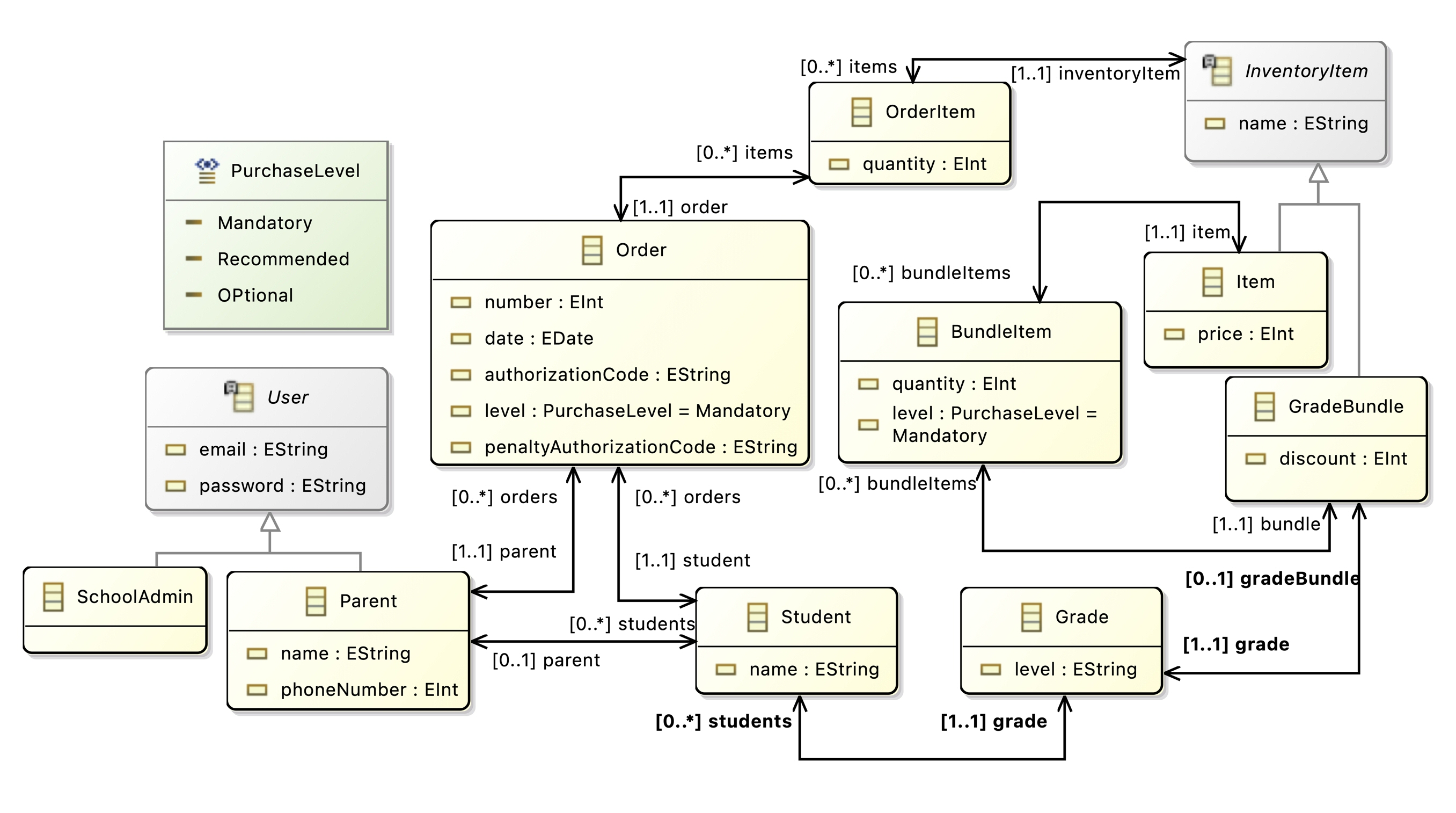}}
    \caption{Expert-developed \emph{CoolSupplies} domain model}\label{fig:domain-model-sample}
\end{figure}


\begin{figure}
    \centering
    \fbox{\includegraphics[width=\linewidth]{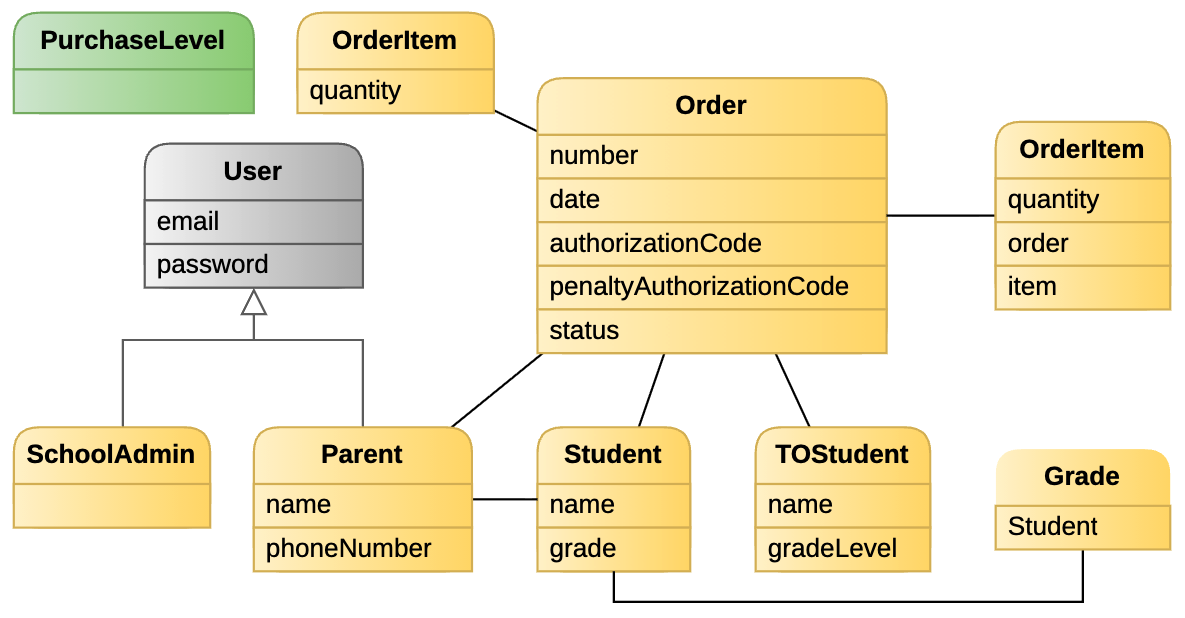}}
    \caption{Reverse engineered \emph{CoolSupplies} model through our pipeline (\textit{excerpt})}
    \label{fig:generated-example}
\end{figure}

\begin{figure*}
    \centering
    \includegraphics[trim={0 .5cm 0 .5cm},clip,width=.75\linewidth]{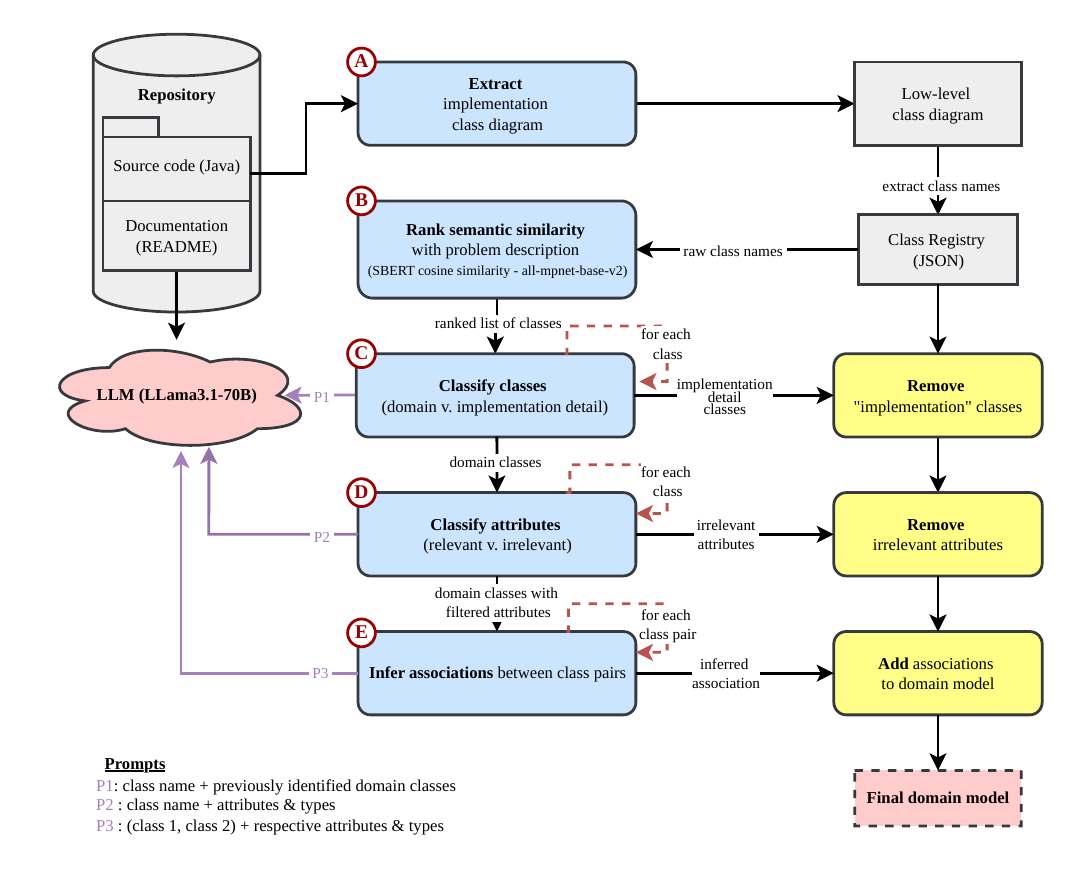}
    \caption{Overview of the Domain Model Extraction Pipeline}\label{fig:extraction-pipeline}
\end{figure*}



Our approach aims to extract domain models from source code using lightweight language models under limited contexts. To achieve this, we adopt a multi-stage and iterative pipeline that refines a domain model from an implementation-level representation.


\noindent\textbf{Illustrative example:} Given the CoolSupplies project, which is a school-oriented system that enables administrators to manage grades, students, and supply bundles, while allowing parents to register, select students, and place orders of required school supplies for a given academic year. Its expert-curated domain model is shown in Figure~\ref{fig:domain-model-sample}, which we use as ground truth for evaluating our approach.
Two independent implementations of this domain model exist. We consider one implementation, comprising 90 classes and 7,899 lines of code, and aim to reverse engineer its source code into the corresponding domain model. Figure~\ref{fig:generated-example} illustrates a corresponding reverse-engineered model (excerpt) produced by our pipeline, and we discuss and analyze the differences between the extracted model and the ground truth model in Section~\ref{sec:evaluation}.\\

\noindent\textbf{Overview:} As illustrated in Figure~\ref{fig:extraction-pipeline}, the process starts by extracting an implementation UML class diagram from the source code \circled{A}, which serves as an initial approximation of the domain model. We then rank the extracted classes according to their likelihood of representing domain concepts using semantic similarity with project documentation \circled{B}. Rather than analyzing the entire system at once, we iteratively process this ranked list, allowing the language model to incrementally build contextual knowledge of the domain despite operating with limited context. The pipeline proceeds through successive refinement steps, including the classification of classes into domain or implementation elements \circled{C}, the filtering of irrelevant attributes \circled{D}, and the inference of relationships between domain concepts \circled{E}. Each step combines structural heuristics with LLM-based reasoning, and leverages previously inferred information to guide subsequent decisions. 

The proposed approach targets object-oriented systems and assumes that an implementation-level class diagram can be extracted from the source code. It requires a short, domain-level description of the system (e.g., a README or requirements description), rather than a rigid documentation format.
Class and attribute identifiers should retain sufficient semantic information to reflect the application domain. These assumptions underpin the semantic ranking and iterative classification stages. The following subsections describe these components as well as the remainder of the pipeline.

\subsection{Documentation and Diagram Extraction}


In this work, we focus on object-oriented systems, using Java projects as a representative setting. 
Additionally, our approach leverages domain documentation from the source code repository. 
This documentation need not be exhaustive or fully up to date, as its primary purpose is to introduce core domain concepts and establish a shared vocabulary.



In step \circled{A}, we use a model-driven static analysis tool to automatically derive from the source code an implementation-level representation capturing the structural elements of the system, including classes, attributes, and relationships. 

This representation serves as the initial view of the system and may include both domain concepts (e.g., an \texttt{Order} class) and implementation artifacts (e.g., a \texttt{CoolSuppliesController} class). To facilitate subsequent processing, it is transformed into a simplified textual class registry that records classes, attributes, relationships, and other relevant structural information while omitting non-essential details such as package hierarchies. Method signatures are retained as context to support semantic abstraction but are omitted from the final domain model, which captures only structural domain concepts.
The registry can be serialized into any machine-readable representation, such as JSON.

Consequently, once such a structural representation is available, the remainder of the pipeline operates exclusively on this intermediate representation together with the available project documentation. Adapting the approach to another object-oriented language therefore primarily requires producing an equivalent structural representation as input to the pipeline



\subsection{Semantic similarity ranking}

A key challenge in our setting is that, at the beginning of the process, the language model has no explicit knowledge of the domain boundaries. As a result, directly classifying classes as domain-related or implementation-related in an arbitrary order may lead to early misclassifications, which can negatively impact subsequent decisions due to the iterative nature of the approach.

To mitigate this issue, we introduce an ordering strategy that increases the likelihood of correct early classifications. Our intuition is that classes representing domain concepts are more likely to be reflected in high-level project documentation, whereas implementation-specific classes are less frequently described. Based on this heuristic, we estimate the semantic relevance of each class with respect to the documentation.

Concretely, in step \circled{B}, we represent both class names and documentation fragments using vector embeddings and compute their semantic similarity using cosine similarity. 
In our running example, a class such as \texttt{SchoolAdmin} receives higher cosine similarity scores ($0.5991$), whereas an implementation detail class like \texttt{XmlSerializer}, which is unrelated to the domain, receives a much lower score ($0.0840$). Hence, each class is ranked according to its similarity with the documentation, and this ranked list is then used to guide the iterative classification process. 
This approach fosters the intuition that classes most likely to represent domain concepts are considered first, allowing the language model to establish an initial and more reliable understanding of the domain, which is progressively refined as additional classes are analyzed. 
Since this ranking relies only on the semantic correspondence between implementation-level identifiers and high-level domain terminology, the concise project description introduced in Step \circled{A} is generally sufficient to guide the process.

However, this ranking alone is not a sufficient heuristic, as some implementation detail classes may also contain domain-related terms, for instance, \texttt{CoolSuppliesFxmlView} scores $0.4888$ despite being primarily tied to the UI implementation. 
To this effect, we augment the workflow using an LLM 
to classify the ranked model elements as either \emph{domain-specific} or \emph{implementation detail} elements, as detailed in the following section. 





\subsection{Classifying classes}
The goal of step \circled{C} is to classify each class as either \emph{domain-specific (DS)} or \emph{implementation detail (ID)}, aiming to transform the implementation-level representation into a domain-oriented abstraction.
Rather than performing this classification independently, we adopt an iterative process in which classes are analyzed sequentially following the order established in the previous step. This allows the model to progressively build contextual knowledge about the domain and improve its performance over time. 

\subsubsection{Iterative classification process.}
The project documentation (included in the classification prompt) provides an initial understanding of the application domain but does not establish how domain concepts are represented in the implementation. Classes are therefore processed sequentially, with previously classified concepts included in subsequent prompts to progressively build this mapping.

This enriched context enables the model to approximate global reasoning while operating under limited context. Furthermore, because classes are processed according to their semantic similarity with the project documentation, likely domain concepts are typically identified first, allowing the domain-specific (DS) context to grow rapidly and improving the reliability of later classifications.


\subsubsection{LLM-based classification strategy.}

At each iteration, the classification decision is performed by a language model that evaluates whether a given class represents a domain concept or an implementation artifact. The decision is based on multiple sources of information: (i) the class name, (ii) the project documentation, and (iii) the set of previously classified classes.

The model is guided by general software engineering heuristics, which are used as soft indicators. In particular, domain classes typically correspond to real-world entities or core concepts of the application domain, while implementation classes often relate to architectural patterns, data access layers, or technical mechanisms. Naming conventions can provide useful signals (e.g., \texttt{Controller}, \texttt{Repository}, \texttt{DTO}), but they are inherently ambiguous
as similar terms may refer to either domain concepts or implementation artifacts depending on the application (e.g., \textit{air traffic controller} vs. MVC \textit{Controller}).

The classification is formulated as a strict binary decision. To support this process, we design structured prompts that provide the model with the relevant context and guide its reasoning through instructions and examples. The prompt includes the project documentation, previously classified classes, and representative examples of domain and implementation elements \cite{replication}.

\subsubsection{Filtering implementation elements.}
Once all classes have been classified, we have domain-specific and implementation-detail classes. The domain model is then constructed by retaining only DS classes and removing ID classes from the initial diagram. Since relationships in the original diagram may involve ID classes, removing them can lead to the loss of connections between domain concepts. This motivates an additional step to preserve meaningful structural relationships. 

\subsubsection{Preserving structural relationships.}

To mitigate the loss of relationships induced by the removal of implementation classes, we introduce a mechanism to recover indirect associations between domain classes. The initial class diagram is interpreted as a graph, where nodes represent classes and edges represent relationships.

We identify paths of the form $D_1 \rightarrow I_1 \rightarrow \dots \rightarrow I_n \rightarrow D_2$, where $D_i$ are domain classes and $I_j$ are implementation classes. For each such path, we introduce a direct association between $D_1$ and $D_2$, effectively bypassing the implementation layer. Existing direct associations between domain classes are preserved.

While this transitive association mechanism helps maintain connectivity and improves recall, it may introduce spurious relationships when implementation classes connect unrelated domain concepts. In practice, we observed that retaining these inferred associations yields better overall results than relying solely on direct relationships.

\subsection{Classifying attributes}

In step \circled{D}, we refine the classes identified in step \circled{C} by retaining only the attributes that contribute to the representation of domain concepts, and filtering out those related to implementation concerns.
For example, an \texttt{OrderItem} class in the implementation codebase may include a \texttt{cachedHashCode} attribute, which reflects a low-level implementation concern (e.g., performance optimization) rather than a domain concept, and should therefore be removed at this stage.
For each class, we consider its attributes and their types as extracted from the initial class diagram. The task is formulated as a binary classification problem, where each attribute is 
either \emph{relevant} or \emph{irrelevant} to the domain entity represented by the class.

Attribute classification is performed iteratively. For a given class, attributes are processed sequentially, and previously classified attributes are used as contextual information to guide subsequent decisions. 
This allows the model to progressively identify patterns within the attributes of a class and improve classification consistency.
In addition, the classification of attributes is contextualized by considering the surrounding structure of the model. In particular, neighboring classes, identified through associations, are provided as context, as some attributes may correspond to relationships rather than intrinsic properties of the class.

\subsubsection{LLM-based classification strategy.}

At each iteration, a language model evaluates whether a given attribute contributes to the domain representation of the class. The decision is based on (i) the class name, (ii) the attribute name and type, (iii) the project documentation, (iv) previously classified attributes, and (v) neighboring classes, all of which are included in the prompt used \cite{replication}.
The model is guided by general modeling heuristics. Relevant attributes typically describe intrinsic properties of domain entities (e.g., identifiers, descriptive fields, or domain-specific characteristics), while irrelevant attributes often correspond to technical concerns such as temporary variables, flags, storage mechanisms, or implementation-specific data structures. 


Once all attributes have been classified, irrelevant attributes are removed from each class. The resulting representation retains only domain-specific classes and their relevant attributes, forming a domain model that abstracts away implementation-level details.

\subsection{Inferring associations between class pairs}

After filtering classes and attributes, the resulting model may still lack meaningful relationships between domain concepts. While some associations are preserved from the original class diagram or recovered through transitive connections, others may be missing due to the abstraction process. Removing implementation elements can break indirect links between domain classes, and some relationships may not be explicitly represented in the code structure.

To address this, we introduce a final step \circled{E} to infer potential associations between domain-specific (DS) classes, with the goal of improving the completeness and coherence of the resulting domain model.
We consider all pairs of DS classes that are not already connected through either direct associations or previously inferred transitive associations. These pairs represent potential candidates for new relationships.
This step reduces the search space to unresolved class pairs, allowing the model to focus on identifying missing connections rather than reconsidering established ones.

\subsubsection{LLM-based association inference.}

For each candidate pair, the language model is used to determine whether a meaningful association should exist between the two classes. The decision is based on (i) the class names, (ii) their attributes, methods and types, (iii) the project documentation, and (iv) existing relationships in the model, which are included in the prompts \cite{replication}.

The model is guided by general modeling heuristics. In particular, associations may be inferred when (i) one class appears as the type of another program element (e.g., an attribute, method parameter, or return value), suggesting an implicit structural relationship, or (ii) the corresponding domain concepts are semantically related according to the documentation. These signals are interpreted jointly to determine whether a direct association should be introduced.
The classification is formulated as a binary decision (\textit{association} vs.\ \emph{no association}). 

The final set of associations is obtained by combining (i) the original associations extracted from the class diagram, (ii) the transitive associations introduced earlier, and (iii) the newly inferred associations. 
The current pipeline recovers associations between domain classes without directionality or multiplicities.

\section{Evaluation}\label{sec:evaluation}

\subsection{Research Questions}

The goal of this evaluation is to assess the effectiveness of our approach for extracting domain models from source code using lightweight, locally deployable language models, as well as to analyze the impact of its key design choices. To this end, we address the following research questions:

\textbf{RQ1}: \emph{What is the overall performance of the proposed approach in extracting domain models from source code?} 
This question evaluates how accurately the approach identifies domain classes, relevant attributes, and associations when compared to a reference domain model.


\textbf{RQ2}: \emph{Does preprocessing class names through tokenization improve the performance of the approach?}
This question investigates whether splitting compound identifiers into meaningful tokens (before step \circled{A} in Figure~\ref{fig:extraction-pipeline}) improves semantic interpretation by the language model. 
For example, \texttt{ClientController} becomes ``client controller'' before embedding and classification, potentially improving classification accuracy.


\textbf{RQ3}: \emph{What is the impact of semantic similarity-based ranking on the performance of the approach?} 
This question evaluates the contribution of the ranking strategy (step \circled{B} in Figure~\ref{fig:extraction-pipeline}) by comparing the proposed approach with variants that process classes without prioritization.

\textbf{RQ4}: \emph{How sensitive is the proposed approach to different implementations of the same domain model?} 
This question evaluates the robustness of the approach by comparing its performance across multiple implementations of the same domain.

\subsection{Experimental Setup}

We present here the dataset, metrics and implementation details for the evaluation of our pipeline. These elements are included in our replication package \cite{replication}.

\subsubsection{Dataset.}



We evaluate our approach on ten publicly available Java software engineering projects developed as part of an undergraduate Model-Driven Engineering course \cite{replication}. In this course, students implement an instructor-provided domain model, resulting in software systems for which the intended conceptual model is known \textit{a priori}.
These projects correspond to seven distinct domain models spanning different application domains, meaning three have two independent implementations. 

The presence of multiple implementations for the same domain model allows us to assess the robustness of the approach across different code bases representing the same conceptual domain. Table~\ref{tab:dataset-caracts} summarizes the main characteristics of the reference domain models used in our evaluation, including the number of classes, attributes, and relationships.

All projects are implemented in Java and follow standard object-oriented design practices, making them suitable for our pipeline, which operates on implementation-level class diagrams extracted from source code. In addition to the source code, the pipeline relies on a short textual description of the system and its application domain. Although the descriptions used in our experiments were curated by domain experts, they are representative of the high-level information typically available in project documentation, such as README files or software requirements specifications.

\begin{table*}[h]
\centering
\caption{Overview of evaluated systems (cases) in our dataset}
\vspace{-.3cm}
\begin{tabular}{|c|l|l|c|c|c|c|c|}
\hline
\textbf{Case} & \textbf{Domain name} & \textbf{Description} & \multicolumn{2}{c|}{\textbf{Domain Model}} & \multicolumn{3}{c|}{\textbf{Implementation}} \\
\cline{4-8}
 &  &  & \#classes & \#constructs & Doc \#words & \#classes & LOC \\
\hline
Case 1 & Flexibook (imp 1) & Service booking & 13 & 45 & 146 & 33 & 7787 \\
\hline
Case 2 & Flexibook (imp 2) & Service booking & 13 & 45 & 146 & 36 & 9348 \\
\hline
Case 3 & Climbsafe (imp 1) & Book climbing activities & 13 & 35 & 160 & 68 & 5763 \\
\hline
Case 4 & Climbsafe (imp 2) & Book climbing activities & 13 & 35 & 160 & 56 & 18615 \\
\hline
Case 5 & BikeTourPlus & Book and manage bike tours & 11 & 33 & 119 & 58 & 15408 \\
\hline
Case 6 & AssetPlus & Hotel asset management & 11 & 28 & 122 & 95 & 12639 \\
\hline
Case 7 & CoolSupplies (imp 1) & Manage school supplies & 10 & 20 & 107 & 95 & 8048 \\
\hline
Case 8 & CoolSupplies (imp 2) & Manage school supplies & 10 & 20 & 107 & 90 & 7899 \\
\hline
Case 9 & CheECSEManager & Manage cheese aging & 13 & 38 & 135 & 122 & 11611 \\
\hline
Case 10 & BTMS & Transportation management & 7 & 12 & 38 & 24 & 3509 \\
\hline
\end{tabular}
\label{tab:dataset-caracts}
\end{table*}

This dataset allows us to evaluate the ability of our approach to recover domain concepts, attributes, and relationships from real-world code bases. However, it also presents some limitations. First, the number of projects is relatively small, which may limit the generalizability of the results. 
Second, conceptual domain models are subjective and may vary in abstraction. The reference models therefore represent one plausible conceptualization rather than a unique ground truth.=
Finally, the projects are of moderate size, and larger industrial systems may introduce additional challenges not fully captured in this evaluation.

\subsubsection{Evaluation Metrics.}

We evaluate the extracted domain models using standard information retrieval metrics: \textbf{precision}, \textbf{recall}, and \textbf{F1-score}. These metrics are automatically computed separately for classes, attributes, and associations.
To account for variability in LLM outputs, each experiment is executed three times, and we report the average precision, recall, and F1-score across runs.

For \textbf{classes}, true positives correspond to correctly identified domain classes, false positives to implementation classes incorrectly classified as domain, and false negatives to missing domain classes.

For \textbf{attributes}, evaluation is restricted to correctly identified domain classes (true positives). An attribute is considered correct if it belongs to a true-positive class and matches an attribute in the reference model. Precision and recall are computed based on the set of attributes retained after filtering.

For \textbf{associations}, evaluation is more complex due to the transformations applied to the initial class diagram. We consider an association to be correct if it connects two domain classes that are also connected in the reference model. Similar to attributes, associations are evaluated with respect to the set of correctly identified domain classes. This avoids penalizing the approach for discrepancies introduced by errors in earlier classification steps. In particular, only associations between true-positive classes are considered when computing precision and recall.
Finally, we report macro-averaged precision, recall, and F1-scores across all projects.

\subsubsection{Implementation Details}


\paragraph{Proposed approach.}
Our approach uses the open-source LLaMA3.1-70B language model, executed locally, across multiple stages of the pipeline, including class classification, attribute filtering, and association inference. The intermediate representation of each repository is extracted using Visual Paradigm's \textit{Instant Reverse} functionality.



\paragraph{Experimental protocol.}
Each experiment is repeated three times to account for variability in LLM outputs, and we report average results. For ablation studies (RQ2 and RQ3), we compare the default configuration of our approach with variants that modify specific components (e.g., tokenization, ranking strategy).

\paragraph{Hardware configuration.}
All experiments are conducted on a machine equipped with 4 NVIDIA GeForce RTX 3090, with 24 GB of GPU memory. The use of locally deployable models is a key aspect of our approach, and this setup reflects realistic hardware constraints in average industrial environments.

\section{Evaluation Results}\label{sec:results}

\subsection{RQ1: Overall Performance}


\paragraph{Recovery performance. }

Table~\ref{tab:full-perf-metrics} reports the aggregated performance of our approach across all ten projects, while Table~\ref{tab:results} provides detailed results for each individual case. The full results along with the scripts used are included in our replication package \cite{replication}.

Overall, the approach achieves strong performance across all elements of the domain model. For \emph{classes}, we obtain a precision of 0.84, a recall of 0.97, and an F1-score of 0.90, indicating that most domain concepts are correctly identified, with limited confusion with implementation elements. For \emph{attributes}, performance is even higher, with an F1-score of 0.94, reflecting the effectiveness of the attribute filtering step once domain classes are correctly identified.
For \emph{associations}, evaluated only between correctly identified domain classes, the approach achieves a precision of 0.78, a recall of 0.93, and an F1-score of 0.85. While recall remains high, precision is lower compared to classes and attributes, indicating that the approach tends to infer additional relationships that are not always present in the reference model. 
Since associations are inferred only between classes retained in the extracted domain model, evaluating associations involving false-positive or false-negative classes would conflate errors originating from different stages of the pipeline. Restricting the evaluation to true-positive classes therefore isolates the quality of the association recovery process itself.


\paragraph{Runtime}

The end-to-end execution time is approximately 30 minutes on our hardware configuration and remained relatively stable across the evaluated projects, which range from 24 to 122 implementation classes. Since domain model reverse engineering is primarily an offline program task, i.e., batch processing, rather than a runtime activity, this execution time is aligned with the intended use cases~\cite{yoon2020log}. 

\begin{table}[h]
    \centering
    \caption{Average precision, recall, and F1-score across all cases. }
    \vspace{-.3cm}
    \begin{tabular}{|l|ccc|}
    \hline
         & Precision & Recall & F1 \\
        \hline
        Classes & 0.84 & 0.97 & 0.90 \\
        \hline
        Attributes & 0.89 & 0.99 & 0.94 \\
        \hline
        Associations (TP classes only) & 0.78 & 0.93 & 0.85 \\
        \hline
    \end{tabular}
    
    \label{tab:full-perf-metrics}
\end{table}

The detailed results in Table~\ref{tab:results} show that recall is consistently high across all projects, often close to or equal to 1.00 for classes and attributes. This suggests that the approach is effective at capturing domain elements and rarely misses relevant concepts. In contrast, precision varies more across cases, particularly for associations, where inferred relationships may introduce noise.

Finally, performance differences across projects appear to be more influenced by the nature of the application domain than by model size or complexity. While the reference domain models are of comparable size and structure (Table~\ref{tab:dataset-caracts}), the corresponding implementations vary more significantly in terms of number of classes and lines of code (Table~\ref{tab:dataset-caracts}). Despite this variability, no clear correlation is observed between implementation size and performance. Instead, domains with clearly defined entities and relationships tend to yield higher precision. For instance, systems such as \emph{CoolSupplies} and \emph{CheECSEManager}, which involve concrete entities (e.g., supplies, orders, or cheese batches) and explicit relationships between them, achieve higher precision, particularly for associations. In contrast, domains involving more implicit or abstract relationships, such as \emph{Climbsafe} or \emph{AssetPlus}, lead to increased ambiguity. In these cases, concepts such as bookings, assets, or operational processes can be represented in multiple ways, making it more difficult for the model to distinguish between domain and implementation elements and to infer precise relationships.

\textbf{Answer to RQ1:} The proposed approach achieves high recall and strong overall performance in extracting domain models from source code, with particularly strong results for classes and attributes. While association inference introduces some noise, the approach remains effective across diverse projects.

\begin{table}[t]
\centering
\caption{Performance of the proposed approach across all cases (averaged over 3 runs)}
\vspace{-.3cm}
\begin{tabular}{|c|ccc|ccc|ccc|}
\hline
\textbf{C.} & \multicolumn{3}{c|}{\textbf{Classes}} & \multicolumn{3}{c|}{\textbf{Attributes}} & \multicolumn{3}{c|}{\textbf{Associations}} \\
\cline{2-10}
 & P & R & F1 & P & R & F1 & P & R & F1 \\
\hline
1 & 0.85 & 1 & 0.92 & 0.90 & 1 & 0.95 & 0.72 & 0.94 & 0.81 \\
\hline
2 & 0.82 & 1 & 0.90 & 0.88 & 1 & 0.94 & 0.70 & 0.91 & 0.79 \\
\hline
3 & 0.78 & 0.92 & 0.84 & 0.85 & 0.96 & 0.90 & 0.43 & 0.89 & 0.58 \\
\hline
4 & 0.80 & 0.92 & 0.86 & 0.87 & 0.96 & 0.91 & 0.45 & 0.90 & 0.60 \\
\hline
5 & 0.88 & 1 & 0.94 & 0.92 & 1 & 0.96 & 0.75 & 0.92 & 0.83 \\
\hline
6 & 0.76 & 0.91 & 0.83 & 0.82 & 0.95 & 0.88 & 0.47 & 0.88 & 0.61 \\
\hline
7 & 0.90 & 1 & 0.95 & 0.93 & 1 & 0.96 & 0.86 & 0.93 & 0.89 \\
\hline
8 & 0.89 & 1 & 0.94 & 0.92 & 1 & 0.96 & 0.84 & 0.92 & 0.88 \\
\hline
9 & 0.87 & 1 & 0.93 & 0.91 & 1 & 0.95 & 0.79 & 0.94 & 0.86 \\
\hline
10 & 0.86 & 1 & 0.92 & 0.90 & 1 & 0.95 & 0.78 & 0.93 & 0.85 \\
\hline
\end{tabular}
\label{tab:results}
\end{table}





\begin{table}
    \centering
    \caption{Evaluation of the impact of removing semantic similarity ranking and the contribution of class name tokenization. The default configuration (\xmark, \checkmark) corresponds to the results reported in Table~\ref{tab:full-perf-metrics}.}
    \vspace{-.3cm}
    \begin{tabular}{|c|c|ccc|}
    \hline
        Tokenization & Ranking & Precision & Recall & F1 \\
        \hline
        \xmark & \xmark & 0.55 & 0.95 & 0.69 \\
        \hline
        \xmark & \checkmark & \textbf{0.84} & \textbf{0.97} & \textbf{0.90} \\
        \hline
        \checkmark & \xmark & 0.55 & 0.95 & 0.70 \\
        \hline
        \checkmark & \checkmark & 0.50 & 0.90 & 0.64 \\
        \hline
    \end{tabular}
    \label{tab:abl-study-results}
\end{table}

\subsection{RQ2: Impact of Tokenization} 
To evaluate the impact of class name tokenization, we compare the default configuration of the approach (no tokenization) with its tokenized variant under the same conditions (Table~\ref{tab:abl-study-results}). 

The results show that tokenization does not improve performance across any of the evaluated metrics. Compared to the default configuration, tokenization leads to a decrease in precision (from 0.84 to 0.50), recall (from 0.97 to 0.90), and consequently F1-score (from 0.90 to 0.64). This indicates that tokenization not only fails to provide benefits, but can significantly degrade performance.

A likely explanation is that splitting compound identifiers removes useful semantic and structural information embedded in class names. Modern language models are able to interpret such identifiers effectively without preprocessing, and tokenization may disrupt these representations.

\textbf{Answer to RQ2:} Tokenization does not improve performance and can significantly degrade it. The default approach without tokenization is therefore preferable. 

\subsection{RQ3: Impact of Semantic Similarity Ranking}

To evaluate the impact of semantic similarity-based ranking heuristic, we compare the default configuration of the approach (with ranking) to a variant that processes classes without prioritization (Table~\ref{tab:abl-study-results}).

The results show that ranking has a significant impact on performance. When ranking is disabled, the F1-score drops from 0.90 to 0.69, with a corresponding decrease in precision (from 0.84 to 0.55), while recall remains relatively high. This indicates that, without ranking, the approach tends to misclassify more implementation elements as domain concepts, leading to lower precision. We also observe this trend when considering the tokenized configuration.


These results highlight the importance of processing classes in an order that prioritizes likely domain concepts. By guiding the iterative classification process, ranking enables the model to establish a more reliable initial context, which improves subsequent decisions and reduces error propagation.

\textbf{Answer to RQ3:} Semantic similarity-based ranking is a key component of the approach, significantly improving precision and overall performance by stabilizing the iterative classification.

\subsection{RQ4: Sensitivity to Implementation Variations}

To assess the robustness of the proposed approach, we compare its performance across projects that share the same domain model but differ in their implementations. In our dataset, three domain models (Flexibook, Climbsafe, and CoolSupplies) are each associated with two independent implementations, allowing us to evaluate the sensitivity of the approach to variations in code structure.

Performance remains consistent across different implementations of the same domain. For each of these domains, the corresponding implementations yield similar precision, recall, and F1-scores across classes, attributes, and associations. This indicates that variations in code organization, naming, or implementation details have limited impact on the extracted domain models.
These findings suggest that the approach captures domain-level semantics rather than relying on specific implementation patterns. In particular, the use of documentation and iterative reasoning enables the model to focus on domain concepts independently of how they are encoded in the code.

\textbf{Answer to RQ4:} The proposed approach is robust to implementation variations, producing consistent results across different code bases that implement the same domain model.

\section{Related work}
\label{sec:related_work}


\subsection{AI-assisted Modeling}

Several works in recent years have explored the use of AI and machine learning techniques to support modeling tasks. Early approaches relied on heuristics and classical NLP techniques. For instance, \cite{yang2022towards} demonstrated the potential of such methods for domain model extraction, albeit with limited performance and generalization capabilities.
With the advent of LLMs, increasing attention has been devoted to this problem \cite{bamouh2025towards}. Recent work has explored their use for extracting various types of models from textual requirements, including sequence diagrams \cite{ferrari2024model}, activity diagrams \cite{khamsepour2025impact}, use case diagrams \cite{kc2024analysis}, and class diagrams \cite{bamouh2025towards, nguyen2026class}. While these approaches show strong potential, they still face challenges in achieving human-level abstractions.

In this work, we focus on the extraction of domain models in the form of class diagrams, leaving the extension to other types of diagrams for future work. Indeed, prior user studies have demonstrated the potential of LLMs for assisting domain modeling tasks, both for class diagram generation \cite{perez2025empowering} and completion~\cite{chaaben2024Utility}.

\subsection{LLMs for Generating Domain Models from Textual Requirements.}

\citet{chen2023automated} investigated the applicability of GPT-3.5 and GPT-4 for domain modeling using use cases derived from software engineering courses. Their results showed improvements over previous approaches, but also highlighted important limitations. In a follow-up work, \citet{yang2024multi} demonstrated that a multi-step, iterative approach yields better results than a holistic one (i.e., prompting an LLM to generate a complete domain diagram in a single step).
This observation is further supported by \citet{shi2025llms}, who compared different strategies for LLM-based domain modeling on similar datasets. Their results confirm that multi-step generation pipelines outperform single-step approaches. In terms of prompt engineering, they found that few-shot prompting provides the best performance.
Similarly, \citet{reinhartz2025leveraging} investigated the use of LLMs (up to GPT-4) for domain modeling from textual descriptions, comparing multiple strategies. Their findings indicate that decomposing requirements into smaller units can improve requirement coverage, albeit at the risk of increased redundancy.
Finally, \citet{nguyen2026class} recently investigated the use of state-of-the-art LLMs (up to GPT-5) for domain diagram generation from textual requirements, as well as their use as evaluators of such diagrams. Their results showed strong performance on both tasks, although complex domains remained challenging.

Overall, these works highlight the complexity of generating domain models from natural language requirements and show that decomposing the task into smaller subtasks, combined with appropriate prompting strategies, is key to achieving high-quality results. 
However, while prior work focuses on deriving models from textual specifications, we address a complementary and less explored problem: extracting domain models directly from source code.





\subsection{AI and LLMs for Reverse Engineering}\label{sec:related_works_reverse_engineering}

Early approaches for Model-Driven Reverse Engineering relied on heuristics and rule-based transformations to derive model elements from source code. For instance, ontology-driven and model transformation techniques have been used to filter and restructure class diagrams \cite{guizzardi2019ontology, bruneliere2014modisco}, while other works employed machine learning classifiers to condense diagrams into more relevant representations \cite{yang2016condensing}. 
More recently, \citet{capuano2022reverse} proposed a RoBERTa-based approach \cite{liu2019roberta} to generate domain models from source code. Although these approaches demonstrate the feasibility of learning abstractions from code, the resulting models often remain close to implementation-level structures, highlighting the difficulty of capturing higher-level domain concepts.

With the emergence of LLMs, new opportunities have arisen for reverse engineering at higher levels of abstraction \cite{zhang2025empirical}. For example, \citet{siala5348203leveraging} showed that LLMs can accurately extract low-level class diagrams from source code at a level comparable to rule-based tools.
LLMs have also shown promise on more complex tasks, including identifying design patterns in source code \cite{komolov2026design}, extracting natural language requirements \cite{miskell2023automated}, and generating structural (component diagrams) and behavioral (state machine) models from code \cite{hatahet2025generating}.

\paragraph{LLM-assisted Domain Diagram Extraction from Code.}

\citet{campanello2025use} explored the use of LLMs for reverse-engineering domain class diagrams. They compared different single-prompt strategies for generation from source code using proprietary LLMs up to GPT-4, reporting promising results (F1 between 62\% and 75\%) when comparing LLM-extracted domain diagrams with expert-curated ones. Their findings highlight persistent challenges in accurately identifying relevant domain entities and relationships.
A similar approach by \citet{boronat2025mdre} employed smaller proprietary LLMs (\texttt{GPT-4o-Mini} and \texttt{Gemini-1.5-Flash}) to extract domain models from source code. To mitigate context window limitations, they adopted a retrieval-augmented generation (RAG) approach, incorporating an abstract syntax tree (AST) of the source code into the prompt. While their method shows promising results, it is primarily designed to extract diagrams at a lower level of abstraction, closely reflecting implementation details. Our work focuses on identifying which elements should be retained or abstracted away in order to capture only the concepts relevant to the application domain. 


\section{Discussion}






This work represents an initial step towards automated reverse engineering of domain models from source code using lightweight LLMs. While the results are encouraging, several limitations also highlight opportunities for future improvements.


\subsection{Threats to validity}

\paragraph{Failure Modes and Input Representation.}
Errors primarily arise from ambiguity in code and limited domain signals in the documentation. When class names are unclear or overloaded, or when documentation is sparse, the model may misclassify implementation elements as domain concepts or infer spurious associations. For instance, in Figure~\ref{fig:generated-example}, the class \texttt{TOStudent} is incorrectly identified as a domain class, although it corresponds to an implementation-level artifact.
These limitations stem in part from the information provided to the model, which currently relies mainly on class names, documentation, and structural context. Our study does not explicitly evaluate the impact of naming quality, although we do expect performance to decrease when identifiers contain little semantic information. Consequently, naming conventions constitute only one of several signals considered by the model. Project documentation, structural information, and the contextual knowledge accumulated throughout the iterative process provide complementary sources of evidence that may compensate for moderate naming inconsistencies, although meaningful identifiers are generally expected to improve classification accuracy. In future work, richer representations could be considered, such as 
comments, or the class implementation. Providing more detailed code-level context could help disambiguate borderline cases and improve both classification and association inference. The robustness of the approach can also be evaluated under different naming conventions and varying levels of identifier quality, as well as with different documentation sources, such as requirements specifications, or under limited or missing documentation.

\paragraph{Dataset and Generalizability.}
The evaluation is conducted on Java projects following object-oriented design principles, which aligns with our use of class diagrams. Extending the approach to other programming paradigms, such as functional or dynamically typed languages, remains an open challenge.
Moreover, the dataset consists of moderately sized academic projects with relatively well-structured code and documentation. While this provides a controlled evaluation setting, it does not fully reflect the complexity of industrial systems, where code bases are larger and documentation is often incomplete. Future work should therefore include larger and more diverse datasets, including industrial case studies.

\paragraph{Evaluation and Matching Strategy.}
Our evaluation relies on exact matching between generated elements and reference models. This strict criterion may penalize semantically equivalent but syntactically different elements. While this assumption is reasonable in our dataset, where domain concepts are reflected in class names, it may not hold in more heterogeneous systems.
Future work could investigate more flexible evaluation strategies, including semantic similarity between model elements or graph-based matching techniques. Additionally, extending the approach to identify implicit domain concepts, i.e., concepts not directly mapped to a single class, remains an open challenge.

\paragraph{LLM Variability and Stability.}
Finally, the approach relies on stochastic language model outputs. Although we mitigate this variability by averaging results over multiple runs, some fluctuations may still affect individual predictions. Investigating more stable prompting strategies, calibration techniques, or smaller specialized models could further improve robustness and reproducibility.

\subsection{Discussion}

\paragraph{Comparison with existing approaches.}
Alternative approaches exist to improve the context-handling capabilities of LLMs. However, these methods still require high-end hardware to operate effectively. For example, \texttt{Qwen2.5-14B}, which supports contexts of up to 1M tokens \cite{ahmed2025qwen}, can be run on our setup (four NVIDIA GeForce RTX 3090 GPUs, each with 24 GB of memory). 
Nevertheless, we found during our experiments that the available GPU memory is exceeded when the input exceeds 271K tokens, which remains insufficient for handling large industrial codebases. This suggests that large-context open-source LLMs still require hardware resources that are impractical in most real-world settings in order to fully leverage their capabilities. 
In addition, as we discussed in Section \ref{sec:related_work}, existing solutions focus on recovering implementation models or operating at different abstraction levels compared to our approach that recovers domain models from source code.

\paragraph{Automation and human oversight.} Our pipeline operates fully automatically, with limited human intervention at the final validation and refinement of the recovered domain model. However, the iterative design can be improved to support human involvement during execution. By estimating the confidence of each classification, the pipeline can identify uncertain decisions and trigger human intervention only when a confidence threshold is not met. Since later decisions build upon earlier ones, validating these low-confidence classifications before subsequent iterations can improve downstream classifications while allowing users to balance automation and manual oversight.

\subsection{Future work}

\paragraph{Domain Modeling Subjectivity.}
Evaluating domain models is inherently challenging due to the absence of a single ground truth. Multiple valid models may represent the same domain at different levels of abstraction~\cite{mohagheghi2008proof}. As a result, some discrepancies between generated and reference models may reflect alternative modeling choices rather than true errors. For example, the \texttt{status} attribute of the \texttt{Order} class (Figure~\ref{fig:generated-example}) is flagged as incorrect, although it can reasonably be considered part of the domain. This observation suggests that domain model extraction should not be viewed as a fully deterministic task. Instead, future work can explore interactive or configurable approaches, where users guide the level of abstraction or validate intermediate results, enabling the generation of models that better align with specific modeling goals.


\paragraph{Association Multiplicities.}
While our approach successfully retrieves associations between classes, it does not currently address the problem of identifying their multiplicities. Inferring multiplicities is particularly challenging, as this information is often implicit in the codebase and may depend on dispersed implementation details, naming conventions, or domain-specific semantics rather than explicit structural cues. As a result, reliably determining whether an association is one-to-one, one-to-many, or many-to-many requires deeper semantic understanding and contextual reasoning. Future work will investigate extending our approach to this problem by combining heuristics with LLM-based techniques.

\paragraph{Scalability and Context Limitations.}
A central motivation of this work is to enable domain model extraction under limited context constraints. While the proposed iterative approach mitigates this limitation, it introduces a trade-off between locality and global reasoning. Decisions are made based on partial views of the system, which may lead to inconsistencies or suboptimal classifications.
Future work can explore hybrid strategies that combine iterative reasoning with selective context expansion, for example, by dynamically retrieving relevant code fragments or using memory mechanisms to maintain a more global view of the system. Our iterative ranking and classification strategy is introduced to compensate for the limited context available to such models and constitutes the main contribution of the paper.

\section{Conclusion}
\label{sec:conclusion}

In this paper, we present an automated approach for reverse engineering domain models from source code using lightweight, locally deployable large language models. By combining structural and semantic heuristics with iterative LLM-based reasoning, the approach addresses a key limitation of compact models, their restricted context window, by enabling progressive identification of domain concepts and relationships without requiring full-system context. This makes it suitable for privacy-sensitive industrial environments.

Our evaluation shows that the approach achieves competitive performance despite operating under constrained computational resources. These results demonstrate that carefully designed hybrid methods can compensate for the limitations of smaller models and provide a practical alternative to large proprietary systems.

This work highlights the continued relevance of Model-Driven Engineering in the era of AI-assisted software engineering. While advances in LLMs may shift attention away from explicit modeling, 
AI can support the reconstruction, evolution, and use of domain models, reinforcing their role in the software development lifecycle.

Finally, the proposed approach opens several opportunities. It can support educational settings by lowering the barrier to modeling, facilitating downstream tasks such as test generation through richer domain representations, and improving human–AI collaboration by making domain knowledge more explicit and structured. More broadly, automating parts of the modeling process has the potential to reduce effort and improve consistency, contributing to more efficient modeling practices.

\bibliographystyle{ACM-Reference-Format}
\bibliography{references}

\end{document}